\documentclass{article}
\usepackage{spconf,amsmath,graphicx}
\usepackage{enumitem}
\usepackage{spconf,amsmath,epsfig}
\usepackage{url}
\usepackage{subfig}
\usepackage{graphicx}
\usepackage{algorithm,algpseudocode}
\usepackage{amssymb}
\usepackage{color}
\DeclareMathOperator*{\argmin}{argmin} 

\title{Perceptual representations of structural information in images: application to quality assessment of synthesized view in FTV scenario }
\twoauthors
 {Suiyi Ling,  Jing Li, Patrick Le Callet}
	{IPI/LS2N Lab, University of Nantes, France}
	{Junle Wang}
	{Turing Lab, Tencent, China}

\begin{document}
%
\maketitle
\begin{abstract}
As the immersive multimedia techniques like Free-viewpoint TV (FTV)  develop at an astonishing rate, user's demand for high-quality immersive contents increases dramatically. Unlike traditional uniform artifacts, the distortions within immersive contents could be non-uniform structure-related and thus are challenging for commonly used quality metrics. Recent studies have demonstrated that the representation of visual features can be extracted from multiple levels of the hierarchy. Inspired by the hierarchical representation mechanism in the human visual system (HVS), in this paper, we explore to adopt structural representations to quantitatively measure the impact of such structure-related distortion on perceived quality in FTV scenario. More specifically, a bio-inspired full reference image quality metric is proposed based on 1) low-level \textbf{contour} descriptor; 2) mid-level contour \textbf{category} descriptor; and 3) task-oriented \textbf{non-natural structure} descriptor. The experimental results show that the proposed model outperforms significantly the state-of-the-art metrics.
\end{abstract}
\section{Introduction}
\label{sec:intro}
With the rise of 3D displays, head-mounted displays and other advanced display techniques, immersive media applications such as FTV, 3DTV, Virtual Reality (VR) and  LightField (LF) have become a hot topic for media ecosystems. The development of immersive media largely relies on the usage of computer vision/image processing techniques to generate synthetic contents that are likely subject to affect structures of images/videos and the viewing experience, a typical example is the synthesized virtual views in FTV scenario due to the limited camera setting/bandwidth. Quality control of the entire immersive system is thus vital for delivering acceptable quality service to users. So far, the structure-related distortions are challenging for commonly used quality metrics to quantify as they distribute locally and non-uniformly throughout the image/video. One of the best instinctive ways to predict the impacts of the non-uniform structure-related distortions on perceived quality is to employ the representation mechanism within HVS~\cite{ling2018role}.

\begin{figure}[!htbp]
 \includegraphics[width=1\columnwidth]{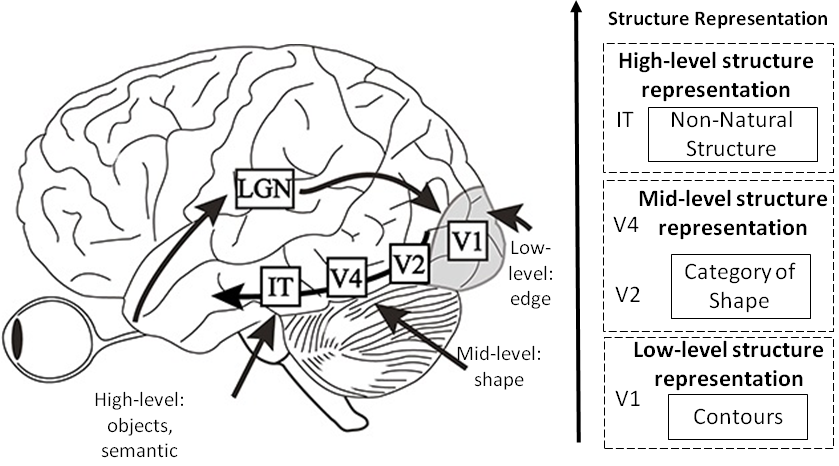}
  \caption{ Left: The hierarchical feedforward framework of visual perception (figure adapted from \cite{manassi2013crowding}). The visual precept is formed based on successive extractions and representations of low-, mid- and high-level features processed through LGN, V1, V2, V4 and IT~\cite{andrews2015low,groen2017contributions,manassi2013crowding}. Right: The proposed hierarchical model for structure-related distortion representation.  }
  \label{fig:perception_basis}
\end{figure}

The process of human analyzing a visual scene has been characterized by the presence of regions in the extrastriate cortex that are selectively responsive to scenes~\cite{groen2017contributions,andrews2015low}. These regions have often been interpreted to represent high-level properties of scenes and they also exhibit substantial sensitivity to low and mid-level properties. A recent bio-vision study~\cite{peirce2015understanding} proposes a hierarchical framework of visual perception, which comprises a series of discrete stages that successively produce three levels of representations. This framework is illustrated in the left part of Figure~\ref{fig:perception_basis}. 

Inspired by the aforementioned bio-vision theories, in this paper, a hierarchical structure representation model is proposed and applied to quality assessment for synthesized images in FTV scenario.  This model consists of three level representations as depicted in the right part of Figure~\ref{fig:perception_basis}, where low-level structure representation of images is defined as local  basic structure information (e.g., local contours); mid-level is defined as intermediate `pattern-based encoded feature', where the patterns are learned by summarizing semantic characteristics of local structure information (e.g., categories of contours); high-level is defined as `task-related abstraction', which learns a set of meaningful abstract structure-related patterns reflecting the characteristics of the task (e.g., non-natural structure in image quality assessment). 

The paper has the following organization: Section~\ref{sec:RW} summarizes the existing quality metrics designed for multi-view images. The proposed hierarchical metric is introduced in Section~\ref{sec:model}. The performance of the proposed metric is reported and analyzed in Section~\ref{sec:ER}. Conclusions are presented in Section~\ref{sec:con}.

\section{Related Work}
\label{sec:RW}
In order to better evaluate the quality of synthesized views in the case of FTV, some metrics are proposed. The very first metric VSQA~\cite{conze2012objective} was proposed using three visibility maps which characterize complexity in terms of textures, diversity of gradient orientations and presence of high contrast. The 3DswIM was introduced by Battisti~\textit{et al.}~\cite{battisti2015objective} based on statistical features of wavelet sub-bands. Stankovi{\'c}~\cite{sandic2015dibr}~\textit{et al.} first deployed morphological wavelet decomposition for quality assessment of synthesized images named MW-PSNR. Later, another metric devises PSNR with morphological pyramids decomposition (MP-PSNR) was proposed in~\cite{sandic2015dibrMP}. Based on the fact that PSNR is more consistent with human judgment when calculated in higher morphological decomposition scales, they further proposed the reduced versions of the two metrics~\cite{sandic2016dibr}, i.e., MW-PSNR$_r$ and MP-PSNR$_r$, which provide better performance compared to the full versions. Targeting the problem that global shifting artifacts are generally over-penalized by point-wise metrics, CT-IQM~\cite{CT_IQA} was proposed using an encoding scheme based on the context tree. To quantify the deformation of curves in synthesized views, an elastic metric based EM-IQM is proposed in~\cite{ling2017EM}. Li~\textit{et al.}~\cite{li2018quality} proposed LOGs by considering both the geometric distortions as well as the sharpness of the images. Apart from the full reference metrics, several no reference metrics are also proposed by the community. In~\cite{tian2017niqsv}, NIQSV was proposed by hypothesizing that high-quality images are consist of flat areas separated by edges. It is then extended to NIQSV+~\cite{tian2018niqsv+} by considering the existence of the dis-occluded regions. Recently, a novel no reference quality metric for synthesized images namely APT was proposed in~\cite{gu2017model}, where the auto-regression (AR) based local image description is employed. In addition to the metrics mentioned above, we believe that there is still room to improve the performance from a perspective of bio-visual structure representation. Details of our proposal are shown in the following section. 

  \vspace{-10pt}
 \section{The proposed metric}
\label{sec:model}

\begin{figure}[!htbp]
 \includegraphics[width=1\columnwidth]{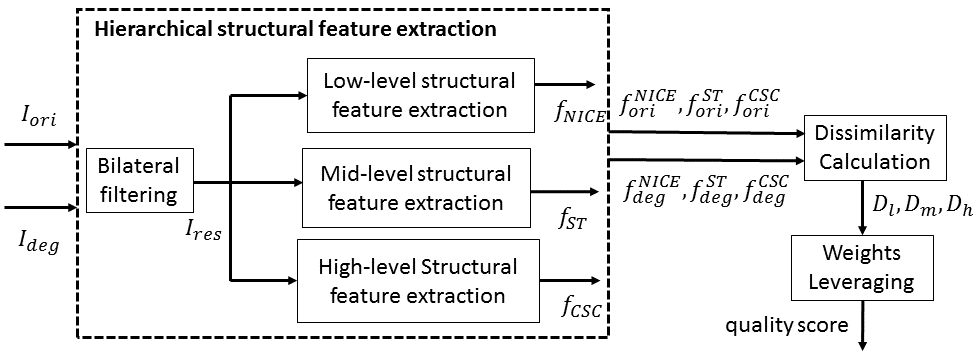}
  \caption{Framework of the proposed hierarchical structural representation based image quality metric for synthesized views. }
  \label{fig:overall_framework}
\end{figure}

In this section, we propose a full-reference image quality metric based on hierarchical structure representation. The proposed framework consists of (1) a pre-processing step for structural information extraction, (2) a hierarchical feature extraction for low, mid and high-level perceptual information extraction, and (3) a pooling step for overall quality score prediction. The overall framework is shown in Figure~\ref{fig:overall_framework}.

 \vspace{-10pt}
 
\subsection{Structural information extraction}
At the very first step of our model, we propose to separate the structural information from textural information. Previous studies ~\cite{ling2018role, rai2016effect} have demonstrated that structural information plays a significantly major role in perceived quality of synthesized image compared to textural information. In addition, it has been shown that bilateral filter has the capability to emphasize such structural information~\cite{ling2018role, rai2016effect}. In our model, we adopt the approximated bilateral filter proposed in~\cite{bib13} for computation efficiency. Afterwards, the responses of bilateral filter for both the degraded image ($I_{deg}$) and its original one ($I_{ori}$), i.e., $f_{ori}^{NICE}$ and $f_{deg}^{NICE}$ are used as the input of the hierarchical feature extraction step. Details are shown in the following sections.

\subsection{Low-level structure representation based estimator}
As pointed out in~\cite{bib10} fragments of contours are the fundamental low-level structure elements that facilitate the successful identification of semantics in images. In~\cite{ling2018role, rai2016effect}, it has been confirmed that the NICE (contour-based image evaluation~\cite{rouse2009image}) descriptor $f^{NICE}$ plays the greatest roles in quantifying the impact of structural distortions on perceived quality, thus, it is adopted in this paper as the low-level structure representation based estimator, which is defined as
\begin{equation}
\begin{aligned}
 D_{l} &= XOR(  f^{NICE}_{ori} , f^{NICE}_{deg}) \\
&=\frac{\sum\limits^{N_c} XOR(C_{ori} \otimes E_{se}, C_{deg} \otimes E_{se})}{N_c}, \\
\end{aligned}
\end{equation}
where $C_{ori}$ and $C_{deg}$ are the contour map detected from the original and degraded images using Canny edge detector, $XOR(\cdot)$ is the point-wise exclusive-or (XOR) operation, and $N_c$ is the number of contour elements. The contour maps are subjected to morphological dilation operation (denoted as $\otimes$ ) with a $3\times3$ `plus-sign' shaped structuring element $E_{se}$ so that the shapes within the images are probed.　
　
  
\subsection{Mid-level structure representation based estimator}
HVS is very efficient in encoding the properties of stimulus by utilizing available regularities. Those efficient representations would be maximally informative with respect to the actual inputs in the world. In particular, low-level elements that share similar characteristics should be encoded more compactly~\cite{kubilius2014encoding}. A higher semantic and efficient representation of low-level structural elements, i.e., mid-level representation is thus defined, which are the categories of the contours. Based on this assumption, a Sketch-Token based Image Quality Metric (ST-IQM) by checking how the categories of contours change due to structural distortions~\cite{ling2017image} is employed as the mid-level structure representation based estimator in our study (also employed in ~\cite{zhou2019quality} as mid-level descriptor), where contours are first `encoded' as a vector $f^{ST}$ of contour categories likelihood values. The mid-level estimator is calculated as the Minkowski summation of the errors computed based on the mid-level descriptor across the entire image: 
 \vspace{-5pt}
\begin{equation} 
 D_{m}= \frac{ [ \sum \limits^{N_p}  D_{JSD}(f_{ori}^{ST},f_{deg}^{ST})^ \beta   ]^{\frac{1}{\beta}} }{N_p}
 \label{equation Minkowski}
\end{equation}

\noindent where $f_{ori}^{ST}$ and $f_{deg}^{ST}$ are the sketch-token descriptors of pairs of matched pixels from the original image to the degraded one (pixels are first matched using a registration methodology proposed in~\cite{sarvaiya2009image} to avoid over-penalizing acceptable global shifting artifacts). $D_{JSD}(\cdot)$ denotes the Jensen\textendash Shannon divergence function, $N_p$ is the number of pixels contained in the image, and $\beta$ is a parameter corresponds to the $\beta-norm$ defining the $L^\beta$ vector space. In our study, $\beta=4$.
 \vspace{-10pt}
 
\subsection{High-level structure representation based estimator}
It is mentioned in~\cite{foldiak2003sparse} that neural code in the higher-level cortex can be sparse, where each element stands for meaningful characteristics of the world (sparsity is considered as one of the essential principles to sensory representation).  In~\cite{ahar2018sparse}, the process of image quality assessment is also assumed to adhere to such a strategy. For quality assessment tasks, the `abstract' elements within the sparse dictionary could be items that reflect quality. For instance, in the case where structure-related distortions are the dominate artifacts, the items could be a set of non-natural structures. In our work, we employed a Convolutional Sparse Coding (CSC) based representation in~\cite{ling2018no,ling2018learn} as a high-level structure representation based estimator. Details are described below.

 First, with a set of patches $Y$ that contain obvious local structure-related distortions collected from synthesized views, a convolutional dictionary $D_Y$ is first learned with a fast CSC algorithm proposed in~\cite{vsorel2016fast} with the equation below 

 \begin{equation}
 \begin{split}
&  \argmin_{D_Y} \frac{1}{2} \Vert y - \sum_{k=1}^K D_k \circledast  Z_k  \Vert^2  \\
&  \mbox{s.t.}~~ \Vert D_k \Vert^2_2 \leqslant 1,   \\
 \end{split}
 \end{equation}
where $\circledast$ denotes the convolution operation, $y \in Y$ denotes training samples, $ Z_k $ represents sparse feature maps, $D_k$ is the $k_{th}$ convolution kernel and $K$ is the number of kernels within the dictionary. 

 With the learned dictionary, for a given $M \times N$ test image $I$, its sparse representation $Z_I$ could be generated with the trained dictionary:
 \begin{equation}
 \argmin_{Z_I} \frac{1}{2} \Vert I-D_Y\circledast  Z_I\Vert^2  + \alpha \Vert Z_I\Vert_1
 \end{equation}
 where $Z_{I}=[Z_1;\ldots;Z_k;\ldots;Z_K]$ is a $M\times N\times K$ tensor of feature maps for $I$, where each map $Z_k$ is the response of using kernel $D_k$. $\alpha$ is a tunable parameter that could be used to balance the model accuracy and the sparsity of feature maps. Finally, a convolutional sparse coding based high-level feature vector $f^{CSC}$ could be then extracted for any image $I$ with:
\begin{equation}
f^{CSC}=( A(Z_1)  ,..., A(Z_K) ) ,
\end{equation}
where $A(\cdot)$ is defined as
\begin{equation}
\label{equ:act}
A(Z_k)= \frac{\sum_{i=1}^M \sum_{j=1}^N \textbf{1}(Z_k(i,j)>\varepsilon) }{ M\times N}
,\end{equation}

$\textbf{1}(c)$ is an indicator function that equals to 1 if the specified binary clause $c$ is true and 0 otherwise, and $\varepsilon$ is a threshold for selecting activated pixels.  Function $A( \cdot )$ aggregates the number of pixels which are above the threshold $\varepsilon$ in each sparse feature map $Z_k$ corresponding to each kernel $D_k$.  Intuitively, this function counts the number of pixels that are activated by the corresponding kernel. Since the kernels are trained to capture structured-related artifacts, this process could be interpreted as the computation of certain types of relative artifacts in the entire image and thus could be used to indicate perceived quality. Finally, support vector regression (denoted as $SVR(\cdot)$) is used to predict the final quality score with the CSC based feature using 1000 times cross-validation. Here, the model that yields the median performance $m_{med}$ is used to compute the high-level structural dissimilarity $D_{h}$: 
\begin{equation}
D_{h} = | SVR(f_{ori}^{CSC}) - SVR(f_{deg}^{CSC}) |
\end{equation}

\subsection{Quality score prediction}
The quality score $S$ is then predicted with the linear combination of the three-level structural distortions $D_{l}$, $D_{m}$ and $D_{h}$ after normalization so that the dissimilarity values are in a range of [0,1]:
 \begin{equation}
\begin{split}
& S = w_{l} \cdot D_{l} +  w_{m} \cdot D_{m} +  w_{h}  \cdot  D_{h}  \\
& s.t.  \ \  w_{l} + w_{m}  + w_{h} =1, \\
\end{split}
\label{eq:bf_m}
\end{equation}
where $ w_{l}$, $w_{m}$, and $w_{h}$  are the weights used for fine-tuning the roles of the low, mid and high-level structural representation based estimators respectively.  

\section{Experimental Results}
\label{sec:ER}
The performance of the proposed model is evaluated on the IRCCyN/IVC DIBR images database~\cite{bosc2011towards}. Images from this database were obtained from three multi-view video plus depth sequences: `Book Arrival', `Lovebird1' and `Newspaper'. Seven DIBR algorithms processed the three sequences to generate four new virtual views for each of them. The database is composed of 84 synthesized views and 12 original frames extracted from the corresponding sequences along with subjective scores. After calculating the differential Mean Opinion Score (DMOS), the following widely employed criteria are utilized to evaluate the performances of the quality metrics:  Pearson Correlation Coefficient (PCC), Spearman’s rank order Correlation Coefficient (SCC) and Root Mean Squared Error (RMSE). Please note that non-linear mapping between the subjective scores and objective measures~\cite{tian2018niqsv+} is conducted before calculating the PCC, SCC, and RMSE.
 
\subsection{Parameters selection}
\begin{figure}[!htbp]
\centering
\includegraphics[width=1\columnwidth]{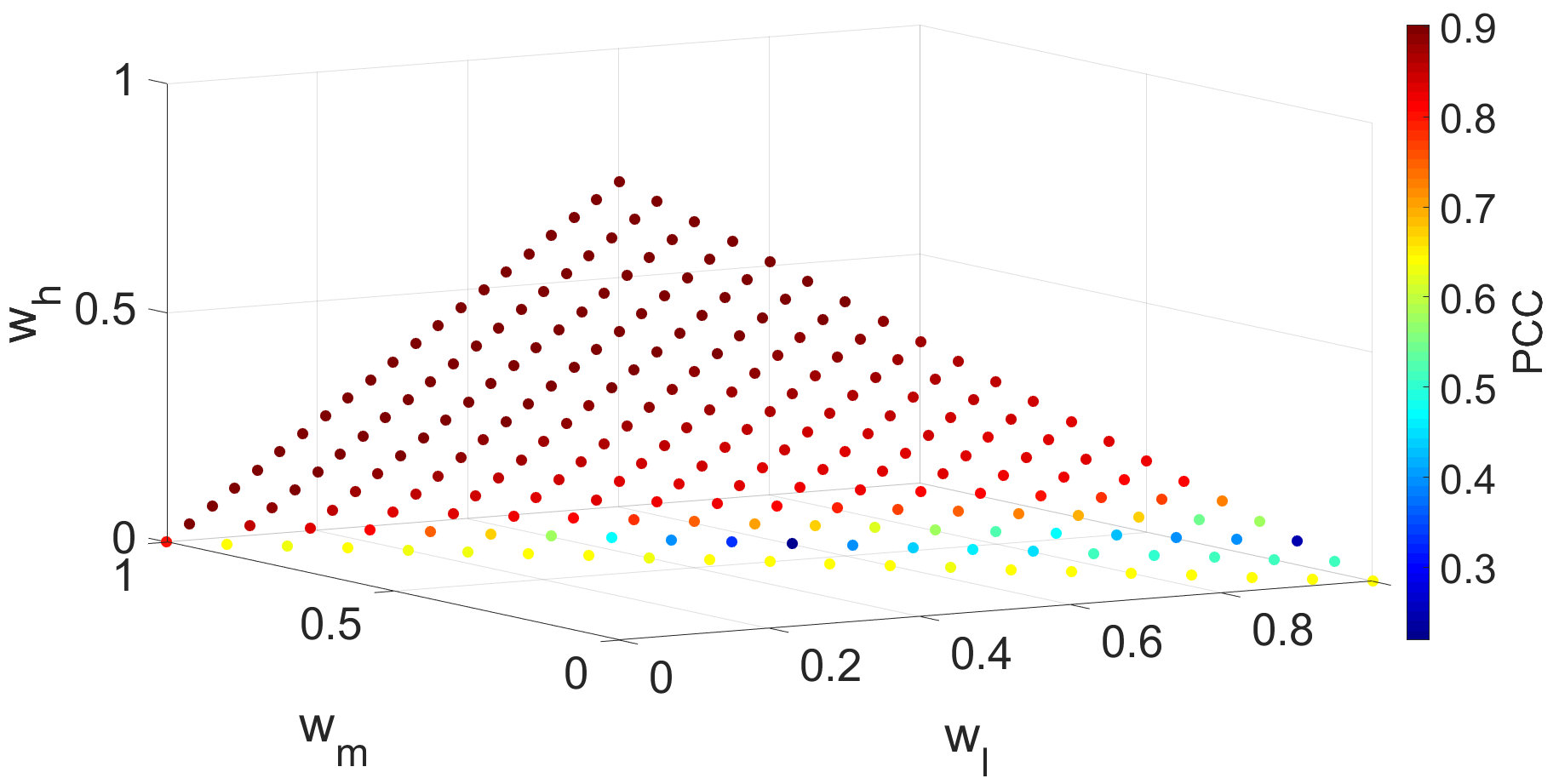}
\caption{Performances of different configurations of $w_{l}$, $w_{m}$ and $w_{h}$. }
 \label{fig:percentage}
\end{figure}

In this study, The overall performance of the proposed model is reported with a configuration of $w_{l}=0.05$, $w_{m}=0.25$ and $w_{h}=0.75$ that obtains the median performance throughout a 1000 cross validation as described in~\cite{ling2019prediction,tian2018niqsv+} with a constraint that the sum of them equals to one. To further analyze the roles of the three levels structural representation in quantifying the impact of structural distortions in predicting quality, the performances of different configurations are checked. The results are shown in Figure~\ref{fig:percentage}. It could be observed from the figure that the performance increases with a higher $w_{h}$. This observation verifies the fact that, high-level structure representation is of greater capability in quantifying structure-related distortions since it is more task-oriented. 
 
 \vspace{-10pt}
 \subsection{Overall performance}
 The overall performance results are shown in Table~\ref{tab:main performance}. According to the table, the proposed hierarchical structural representation based model outperforms the compared state-of-the-art FR/NR image quality metrics designed for quality assessment of synthesized views in FTV scenario. It obtains gains of 9.2\% and 8.6\% in PCC values compared to the second best performing FR metric LoGs and the best performing NR metric CSC-NRM respectively.

\begin{table}[!htbp]
\begin{center}
\caption{\label{tab:main performance}%
Performance comparison of the proposed metric with state-of-the-art metrics }
{
\renewcommand{\baselinestretch}{1}\footnotesize
\begin{tabular}{|c|c|c|c|}
\hline
&\bf{PCC} &\bf{SCC} &\bf{RMSE}      \\ \hline
\multicolumn{4}{ |c| }{ Full Reference Metric (FR)}\\ \hline
3DSwIM~\cite{battisti2015objective} & 0.6864& 0.4842& 0.6125 \\ \hline
MP-PSNR$_{r}$~\cite{sandic2016dibr}& 0.6954& 0.4784& 0.6606  \\ \hline
MW-PSNR$_{r}$~\cite{sandic2016dibr}& 0.6637 &0.4921& 0.6293  \\ \hline
CT-IQM~\cite{CT_IQA} & 0.6809 & 0.6626 &  0.4877  \\ \hline
BF-M~\cite{ling2018role} & 0.6980 & 0.5885 &  0.4768 \\ \hline
EM-IQM~\cite{ling2017EM}& 0.7430 & 0.6726 & 0.4455  \\ \hline
ST-IQM~\cite{ling2017image}&  0.8217 & 0.7710  &  0.3929   \\ \hline
LoGs~\cite{li2018quality} & 0.8256 & 0.7812 & 0.3601 \\ \hline
\bf{Proposed} & \bf{0.9023}&\bf{0.8448} &\bf{0.2870} \\ \hline

\multicolumn{4}{ |c| }{NO Reference Metric (NR)}\\ \hline
NIQSV~\cite{tian2017niqsv}  &0.6346& 0.5146 &0.6167 \\\hline
NIQSV+~\cite{tian2018niqsv+} &0.7114 & 0.4679 & 0.6668\\\hline
APT~\cite{gu2017model} &  0.7307 &  0.7140 &  0.4622  \\\hline
CSC-NRM~\cite{foldiak2003sparse}&   \bf{0.8302}  &  \bf{0.7827}  & \bf{0.3233}   \\\hline
\end{tabular}}
\end{center}
\end{table}

 To analyze if the performances of the proposed metric and other well performed FR and NR metrics are significant, the F-test based on the residual difference between the predicted objective scores and the subjective DMOS values as described in~\cite{liu2015subjective} is employed.  The result is reported in Table~\ref{tab:sinificance_test}, where `1' indicates the quality metric in the row outperforms significantly the one in the column. Thus, the proposed metric outperforms the others significantly. 
 
  \vspace{-10pt}
\begin{table}[!htbp]
\begin{center}
\caption{\label{tab:sinificance_test}%
Statistic significance results based on F-test.  }
{
\renewcommand{\baselinestretch}{1}\footnotesize
\begin{tabular}{|c|c|c|c|c|c|}
\hline
Metric   & LoGs &  ST-IQM & NIQSV+ & APT & CSC-NRM \\\hline
Proposed  & 1 &  1 & 1 & 1 & 1 \\\hline
\end{tabular}}
\end{center}
\end{table}
  \vspace{-10pt}
  \vspace{-10pt}
\section{Conclusion}
\label{sec:con}
Local, non-uniform structure-related distortions within immersive multimedia are challenging for traditional quality metrics.  Inspired by the hierarchical framework of visual perception, in this paper, a 3-level structure representation based model is proposed. This model quantifies the structure-related distortion by checking 1) how local contours change (low-level); 2) how the categories of contour change (mid-level); 3) and the amount of non-natural structure within the synthetic image compared to the original image (high-level). The role of each level of representations on image quality assessment has been studied as well. According to experimental results, the proposed model is significantly superior to the state-of-the-art metrics.


\bibliographystyle{IEEEbib}
{\footnotesize \bibliography{icip_2019.bbl}}

\end{document}